\documentstyle[aps,preprint,pre]{revtex}
\begin{document}

\title{Structure of wavefunctions in (1+2)-body random matrix ensembles}
\author{{\bf V.K.B. Kota$^{1}$ and R. Sahu$^{1,2}$ } \\
$^{1}$ Physical Research Laboratory, Ahmedabad \,\,380 009, India \\
$^{2}$ Physics Department, Berhampur University, Berhampur\,\, 760 007, India}

\maketitle
\begin{abstract}

Random matrix ensembles defined by a mean-field one-body plus a chaos
generating random two-body interaction (called  embedded ensembles of
(1+2)-body interactions)  predict for wavefunctions, in the chaotic domain,
an essentially one parameter Gaussian forms for the energy dependence of  the
number of principal components NPC and the localization length
${\mbox{\boldmath $l$}}_H$ (defined by  information entropy), which are two
important measures of chaos in finite interacting many particle systems. 
Numerical embedded ensemble calculations and nuclear shell model results, for
NPC and ${\mbox{\boldmath $l$}}_H$, are compared with the theory.  These
analysis  clearly point out that for realistic finite interacting  many
particle systems, in the chaotic domain, wavefunction structure is given by
(1+2)-body embedded random matrix ensembles.

\end{abstract}

\pacs{05.45.Mt, 05.30.-d, 21.60.CS, 24.60.Lz}
\date{today}

\def\pr{\prime}
\def\ve{\varepsilon}
\def\be{\begin{equation}}
\def\lan{\left\langle}
\def\ran{\right\rangle}
\def\ee{\end{equation}}
\def\barr{\begin{array}}
\def\earr{\end{array}}
\def\non{\nonumber}
\def\nn8{\nonumber\\[10pt]}
\def\l{\left}
\def\r{\right}
\def\dis{\displaystyle}
\def\ed{\end{document}}
\def\bh{{\mbox{\boldmath $h$}}}
\def\bv{{\mbox{\boldmath $V$}}}
\def\bl{{\mbox{\boldmath $l$}}}
\def\ba{{\mbox{\boldmath $\alpha$}}}
\def\bb{{\mbox{\boldmath $\beta$}}}
\def\bgg{{\mbox{\boldmath $\Gamma$}}}
\def\cod{{\cal O}^\dagger}
\def\co{{\cal O}}
\def\ce{{\cal E}}
\def\cf{{\cal F}}
\def\cg{{\cal G}}
\oddsidemargin 0.0in \evensidemargin 0.5in
\marginparwidth 40pt \marginparsep 10pt
\topmargin 0pt \headsep .5in
\textheight 8.6in \textwidth 6in
\brokenpenalty=10000
\parindent 0.15in

\section{Introduction}

In the last few years the study of quantum chaos in isolated finite
interacting particle systems has shifted from spectral statistics to
properties of wavefunctions and transition strengths (for example,
electromagnetic and Gamow-Teller transition strengths in atomic nuclei,
dipole strengths in atoms etc.). Working in this direction, several research
groups have recognized recently that the two-body random matrix ensembles and
their various extended versions form good models for understanding various
aspects of chaos in  interacting particle systems \cite{Ko-00}.  In
particular using the so called EGOE(1+2), embedded Gaussian orthogonal
ensemble of (1+2)-body interactions defined by a mean-field one-body plus a
chaos generating random two-body interaction, there are now several studies
on the nature of occupancies of single particle states, strength functions
(or local density of states), information entropy, transition strength sums 
and transition matrix elements of one-body transition operators, Fock-space
localization etc., in the chaotic domain of interacting particle systems such
as  atoms \cite{Fl-99}, nuclei \cite{Ko-00,Ze-96},  quantum dots
\cite{Al-97},  quantum computers \cite{Ge-00} and so on. Ref. \cite{Ko-00}
gives a recent overview of this subject. The focus in the present article is
on two important measures of localization (in wavefunctions and transition
strength distributions): (i) number of principal components NPC (or the
inverse participation ratio IPR); (ii) localization length $\bl_H$ as defined
by the information entropy (S$^{info}$). It is well established that the NPC
in wavefunctions characterizes various  layers of chaos in interacting
particle systems \cite{Si-98}. In addition, for systems such as atomic
nuclei, NPC for transition strengths is a measure of fluctuations in
transition strength sums.  Similarly the role of $\bl_H$ in quantum chaos
studies is well emphasized by Izrailev \cite{Iz-90} and more significantly,
using nuclear physics examples \cite{Ze-96a} it is well demonstrated that the
wavefunction entropy S$^{info}$ coincides with the thermodynamic entropy for
many particle systems with two-body interactions of sufficient strength but
only in the presence of a mean-field, i.e. in the chaotic domain but with a
mean-field - therefore the significance of  EGOE(1+2). Clearly deriving the
predictions of EGOE(1+2) for NPC and $\bl_H$ are of considerable importance.
This problem was addressed in Refs. \cite{Gs-97,Ko-98}. In \cite{Gs-97}
results for NPC in wavefunctions, in the so called Breit-Wigner (BW) domain,
are derived. On the other hand, in \cite{Ko-98} results in the so called
Gaussian domain (the BW and Gaussian domains are defined in Sect. II ahead)
are derived  for NPC and $\bl_H$ in transition strength distributions with
only the final results mentioned for wavefunctions. The purpose of the
present paper is to give a detailed derivation of the results mentioned in
\cite{Ko-98} for NPC and $\bl_H$ in wavefunctions  and subject them to
numerical tests. Now we will give a preview.

Section II gives  some of the  basic results for EGOE(1+2).  In Section III,
formulas for NPC and $\bl_H$ in wavefunctions are derived by exploiting the
Gaussian nature and the associated properties of strength functions in
EGOE(1+2). Numerical tests of the  theory are given in Section IV. Finally 
Section V gives concluding remarks.

\section{Basic results for (1+2)-body random matrix ensembles} 

Given $m$ fermions in $N$ single particle states,  assuming at the outset
that the many particle spaces are direct product spaces of the single
particle states, two-body random matrix ensembles (usually called TBRE) are
generated by defining the hamiltonian $H$, which is $2$-body, to be a random
matrix in the $2$-particle spaces and then propagating it to the
${\tiny{\l(\barr{c} N \\ m \earr \r)}}$ dimensional $m$ - particle spaces by
using their geometry (direct product structure); often one considers a GOE
representation in the $2$-particle spaces and then the TBRE is called
EGOE(2);  see \cite{Ko-00} for more details. For a EGOE(2), with $N >> m >>
2$, the normalized state density $\rho(E)=\lan\delta(H-E)\ran$ takes Gaussian
form and it is defined by its centroid $\epsilon=\lan H \ran$ and variance
$\sigma^2=\lan (H-\epsilon)^2 \ran$. In order to explicitly state that the
state density is generated by the hamiltonian $H$, sometimes $\rho(E)$ is
denoted as $\rho^H(E)$ and similarly $\epsilon$ as $\epsilon_H$ and $\sigma$
as $\sigma_H$. Note that the averages $\lan\;\;\;\ran$ are over the
$m$-particle spaces ; in the nuclear physics examples, they are usually over
the $m$-particle spaces  with fixed angular momentum ($J$) and isospin ($T$)
which are good quantum numbers. Just as the state density, given a transition
operator $\co$, the normalized bivariate strength densities (matrix elements
of $\co$ weighted by the state densities at the initial and final energies)
$\rho_{biv; \co}(E_i,E_f) = \l[\lan \cod \co \ran\r]^{-1} \lan \cod \delta (H
- E_f ) \co \delta (H-E_i) \ran$  take bivariate Gaussian form for EGOE(2)
and it is defined by the centroids ($\epsilon_i$, $\epsilon_f$) and widths
($\sigma_i$, $\sigma_f$) of its two marginals and the bivariate correlation
coefficient which is given by   $\lan \cod \;[(H-\epsilon_f)/\sigma_f] \; \co
\; [(H-\epsilon_i)/\sigma_i] \ran\; /\;\lan\cod \co\ran$.  Thirdly, the level
and strength fluctuations follow GOE. Moreover, with the Gaussian form for
the state densities and bivariate Gaussian form for the strength densities,
the strength sums $\lan E \mid  \cod \co \mid E \ran = \sum_{E^\pr}\, \l|\lan
E^\pr \mid \co \mid E \ran \r|^2$ take the form of ratio of two Gaussians,  $
\lan E \mid \cod \co \mid E \ran = \lan \cod \co\ran \rho_{\cod \co:\cg }(E)
/\rho_\cg(E)$ where $\rho_{\cod \co:\cg}(E)= \lan \cod \co \delta(H-E)\ran$
is defined by its centroid $\epsilon_{\cod \co}=\lan \cod \co H \ran / \lan
\cod \co\ran$ and  variance $\sigma_{\cod \co}^2= \lan \cod \co H^2 \ran /
\lan \cod \co\ran  -\epsilon_{\cod \co}^2$; $\cg$ stands for Gaussian.

Hamiltonians for realistic interacting particle systems contain a mean-field
part (one-body part $h(1)$) and a two-body residual interaction $V(2)$ mixing
the configurations built out of the distribution of particles in the
mean-field single particle states; $h(1)$ is defined by the single particle
energies $\epsilon_i,\;i=1,2,\ldots,N$ and $V(2)$ is defined by its
two-particle matrix elements. Then it is more realistic to use EGOE(1+2),
the  embedded Gaussian orthogonal ensemble of random matrices of (1+2)-body
hamiltonians where $\{H\}=h(1) + \lambda \{V(2)\}$; sometimes it is more
convenient to use $\alpha h(1) + \lambda \{V(2)\}$. Here $\{\;\;\}$ denotes
ensemble, $\lambda$ and $\alpha$ are free parameters and $V(2)$ in the two
particle spaces is a GOE with unit matrix elements variance; note that in
general $h(1)$ need not be fixed nor $V(2)$ a GOE (in this general case, the
ensemble is called EE(1+2); see \cite{Ko-00} for more details). At this stage
it is important to stress that all the EGOE(2) results mentioned before are
indeed applicable to EGOE(1+2) but only in the domain of chaos. Given $(m,N)$
and the average spacing $\Delta$ (generated by $h(1)$) of the  single
particle states (without loss of generality one can put $\Delta=1$) it is
possible to find the critical $\lambda$ value  $\lambda_c$ such that for
$\lambda \geq \lambda_c$  there is onset of chaos (GOE level fluctuations) in
many ($m >> 1$) particle spaces. In fact $\lambda_c$ is of the order of the
spacing between $m$-particle mean-field basis states that are directly 
coupled by the two-body interaction; see the second and third reference in 
\cite{Al-97}. For $\lambda > \lambda_c$, for example, it is well established
that the transition strength sums in EGOE(1+2) follow the EGOE(2) forms;  see
Fig. 1c ahead. Refs. \cite{Ko-00,Ko-99} give many numerical examples  for
this, drawn from EGOE(1+2) and more importantly for atomic nuclei in several
parts of the periodic table (detailed discussion of the nuclear physics
examples is given in the last reference of \cite{Ko-99}). It should be
mentioned that  the Gausian forms of state and transition strength densities
are used in \cite{Ko-98} to derive simple  formulas for NPC and $\bl_H$ in
transition strength distributions.

For deriving formulas for NPC and $\bl_H$ in wavefunctions,  most useful
quantity is the strength function (or local density of states) $F_k(E)$.
Given the mean-field basis states $\l. \l| k \r. \ran$ with energies $E_k=
\lan k \mid H \mid k \ran$, the eigenstates $\l. \l| E \r. \ran$ can be
expanded as $\l. \l| E \r. \ran = \sum_k\;C_k^E\;\l. \l| k \r. \ran$. Then
the strength function $F_k(E)=\lan \delta(H-E)\ran^k=\sum_{E^\prime}
\;\l|C_k^{E^\prime}\r|^2\;\delta(E-E^\prime)$ and therefore it gives
information about the structure of the eigenfunctions. In order to proceed
further, let us say that the $E_k$ energies are generated by a hamiltonian
$H_k$ (the structure of $H_k$ is discussed ahead). With this, it is easy to
identify $F_k(E)$ as a conditional density of the bivariate density
$\rho_{biv}(E,E_k)=\lan \delta(H-E) \delta(H_k-E_k) \ran$. Taking
degeneracies of $E$ and $E_k$ energies into account we have,
$$
\barr{rcl}
\rho_{biv}(E,E_k) & = & \lan \delta(H-E) \delta(H_k-E_k)
\ran \nn8
& = & (1/d)\;\dis\sum_{\alpha \in k, \;\beta \in E}\;
\l|C_{k,\alpha}^{E,\beta}\r|^2 \nn8
& = & (1/d)\;\overline{\l|C_k^E\r|^2} \;
\l[d\; \rho^H(E)\r]\;\l[d\; \rho^{H_k}(E_k)\r] \nn8 
\Longrightarrow & &
\earr
$$
\be
\barr{l}
F_k(E)=\rho_{biv}(E,E_k)/\rho^{H_k}(E_k) \nn8
\overline{\l|C_k^E\r|^2} = \rho_{biv}(E,E_k)/\l[d\;\rho^H(E) \;
\rho^{H_k}(E_k)\r]
\earr
\ee
In (1) $d$ stands for the dimensionality of the $m$ particle spaces and 
$\overline{\l|C_k^E\r|^2}$ is the average of $\l|C_k^E\r|^2$ over all the
degenerate states. Let us now examine the structure of $H_k$ and
$\rho_{biv}(E,E_k)$. Firstly it should be noted that the  two-body
interaction $V(2)$ can be decomposed into two parts $V(2)=V^{[0]} + \bv$  so
that $h(1)+V^{[0]}$ generates the $E_k$ energies (diagonal matrix elements of
$H$ in the $m$-particle mean-field basis states). With $m$ particles in $N$
single particle states, there is a $U(N)$ group and with respect to this
group $V^{[0]}$ contains a scalar part $V^{[0],0}$ (a function of $m$), an
effective ($m$-dependent) one-body (Hartree-Fock like) part  $V^{[0],1}$ and
an irreducible 2-body part $V^{[0],2}$. The $V^{[0],0}+V^{[0],1}$ will add to
$h(1)$ giving an  effective one-body part of $H$; $h(1) \rightarrow
h(1)+V^{[0],0}+V^{[0],1} = \bh$. The important point now being that, with
respect to a $U(N)$ norm, the size of $V^{[0],2}$ is usually
very small compared to
the size of $\bh$ in the $m$-particle spaces. With this, $H=\bh+\bv$ and then
the $H_k$ is nothing but $\bh$. The piece $\bv= V(2)-V^{[0]}$ generates the
widths and other shape parameters of $F_k(E)$. It should be added that with
respect to the $U(N)$ norm $\bh$ and $\bv$ are orthogonal and therefore
$\sigma^2_H=\sigma^2_\bh + \sigma^2_\bv$. Definition of $V^{[0]}$, a brief
discussion of its $U(N)$  decomposition etc. are given in Appendix A. For
EGOE(1+2), it is well known that the widths of $F_k(E)$ are in general
constant; see \cite{Fl-96} and Appendix A. The average variance of 
$F_k(E)$'s is given simply by
$$
\overline{\sigma^2_k} =
\sigma^2_\bv = (d)^{-1}\; \sum_{\alpha \neq \beta}\;\l|\lan \alpha \mid H
\mid \beta \ran\r|^2
$$
where $\alpha$ and $\beta$ are $m$-particle mean-field basis states indices.
The results, (i) the norm of the $V^{[0],2}$ part is negligible and (ii) the
widths of the strength functions  are nearly constant (with little
fluctuations) are well verified in a number of examples; see \cite{Fr-83} and
references in \cite{Ko-00} for many nuclear physics examples. EGOE(1+2)
discussions in the literature tacitly assume that $\bh$ is $h(1)$ and $\bv$
is $V(2)$ and the same is assumed from now on, i.e $H=\bh+\lambda \bv
\rightarrow h(1)+\lambda V(2)$. In addition to (i) and (ii), it is well
verified in a number of numerical calculations that: (iii) $F_k(E)$'s exhibit
a transition from BW to Gaussian form in the chaotic domain defined by
$\lambda >  \lambda_{F_k}$; usually $\lambda_c < \lambda_{F_k}$;  see
\cite{Ko-00,KS-00} for some analytical understanding of this result. The
results (i), (ii) and (iii) clearly imply that the $\rho_{biv}(E,E_k)$ is a
bivariate Gaussian and  this result was first mentioned in \cite{Kk-89}. A
numerical example for the BW to Gaussian transition in strength functions in
EGOE(1+2) is shown in Fig. 1a. In this example $\lambda_{F_k} \sim 0.2$ and
it is much larger than $\lambda_c=0.06$ obtained via the results for the
Dyson-Mehta $\overline{\Delta}_3$ level statistic shown  in Fig. 1b. 
Thus there is onset of GOE fluctuations much before 
the $F_k(E)$'s start taking Gaussian form, i.e.
$\lambda_{F_k} > \lambda_c$. Unlike the case with strength functions (also
transition strength densities; see \cite{Ks-00}), as mentioned before, 
strength sums start following the EGOE(2) form  (i.e.
$\rho_{n_i:\cg}(E)/\rho_\cg(E)$) from $\lambda=\lambda_c$. This is 
demonstrated in Fig. 1c where occupancies $\lan E \mid n_i \mid E \ran$  as a
function of $E$ are shown (they correspond to the strength  sums generated by
single state ($\l.\lan i \r.\ran$)  destruction operators). As mentioned in
the introduction, the nature of NPC (which is the inverse of IPR) in
wavefunctions in the $\lambda_c \leq \lambda < \lambda_{F_k}$ domain where
$F_k(E)$ is of BW form (i.e. in the BW domain) was studied in \cite{Gs-97}
while the present article is  concerned with the $\lambda > \lambda_{F_k}$
domain (i.e. the Gaussian domain) where $F_k(E)$ is of Gaussian form.

\section{EGOE(1+2) results for NPC and $\bl_H$ in wavefunctions}

For EGOE(1+2), in the chaotic domain with $\lambda > \lambda_{F_k}$, one has
from Sect. II the results: (i) $E_k$ are generated by $H_k=h(1)$, therefore
the variance of $\rho^{H_k}(E_k)$ is $\sigma^2_\bh$; (ii) widths of the
strength functions are constant and they are generated by $V(2)$, the average
variance $\overline{\sigma^2_k}=\sigma^2_\bv$; (iii) $F_k(E)$'s are Gaussian
in form; (iv) $F_k(E)$ is a  conditional density of the bivariate Gaussian
$\rho_{biv:\cg}(E,E_k)$. The correlation coefficient $\zeta$ of
$\rho_{biv:\cg}(E,E_k)$ is given by,
\be
\zeta = \dis\frac{\lan (H-\epsilon_H) (H_k-\epsilon_H)\ran}{
\dis\sqrt{\lan (H-\epsilon_H)^2\ran\; \lan (H_k-\epsilon_H)^2
\ran}} = \dis\sqrt{\l(1-\dis\frac{\overline{\sigma^2_k}}{\sigma^2_H}\r)}
\ee 
Note that the centroids of the $E$ and $E_k$ energies are both given by
$\epsilon_H=\lan H \ran$.   In (2) the second equality is obtained by using
the orthogonality between $h(1)$ and $V(2)$ operators. It is immediately seen
that the $\zeta^2$ is nothing but the variance of $E_k$'s (the centroids  of
$F_k(E)$) normalized by the state density variance. The
$\rho_{biv:\cg}(E,E_k)$, which takes  into account the fluctuations in the
centroids of $F_k(E)$ and assumes that the  variances are constant,  is used
to derive formulas for NPC and $\bl_H$ in the wavefunctions (methods of
taking into account variance fluctuations will  be discussed ahead) $\psi_E =
\l.\l| E \r. \ran$ expanded in the mean-field basis defined by the states
$\phi_k = \l.\l| k \r. \ran$. Let us first define NPC and $\bl_H$,
\be 
\barr{rcl}
\l| \l. E \ran \r. & = & \dis\sum_{k} \; C^E_k\;\l| \l. k \ran \r. \nn8
\Longrightarrow & & \nn8 
{\mbox{(NPC)}}_E &=& \l[\dis\sum_k\; \l|C^E_k\r|^4 \r]^{-1}
\;\;,\nn8 
\bl_H(E) & = & exp\l[(\mbox{S}^{info})_E\r] /(0.48\;d)\;\;; \nn8 
(\mbox{S}^{info})_E &=& - \dis\sum_{k}\;  \l|C^E_k\r|^2 \;
\ln\;\l|C^E_k\r|^2\;\;. 
\earr 
\ee 
In (3) $0.48d$ is the GOE value for $\mbox{S}^{info}$, thus $\bl_H=1$ for
GOE. Similarly NPC is $d/3$ for GOE.

In terms of the locally renarmalized amplitudes ${\cal C}^E_k=C^E_k/\sqrt{
\overline{\l|C^E_k\r|^2}}$ where the bar denotes ensemble average with
respect to EGOE(1+2), $\dis\sum_k\; \l|C^E_k\r|^4 = \dis\sum_k\; \l|{\cal
C}^E_k\r|^4 \; \l( \overline{\l|C^E_k\r|^2} \r)^2$.  Then the ensemble
averaged (NPC)$_E$ is obtained as follows,
\be
\barr{rcl}
\overline{\dis\sum_k\; \l|C^E_k\r|^4} & \stackrel{EGOE(1+2)}{\longrightarrow}
& \dis\sum_k\; \overline{
\l|{\cal C}^E_k\r|^4} \; \l(\overline{\l|C^E_k\r|^2} \r)^2 \nn8
& = & 3\;\dis\sum_k\;\l(\overline{\l|C^E_k\r|^2} \r)^2 \nn8
& = & \dis\frac{(3/d)}{\l[\rho_\cg^H(E)\r]^2} \dis\int dE_k \;
\dis\frac{\l[\rho_{biv:\cg}(E,E_k)\r]^2}{\rho_\cg^{H_k}(E_k)} \; = \;
\dis\frac{(3/d)}{\l[\rho_\cg^H(E)\r]^2} \dis\int dE_k \;
\rho_\cg^{H_k}(E_k)\,\l[F_{k:\cg}(E)\r]^2 \nn8
{\Longrightarrow} & & \nn8
(\mbox{NPC})_E & = & (d/3)\;\dis\sqrt{1-\zeta^4}\;
exp-\l\{\dis\frac{\zeta^2\;{\hat E}^2}{1+\zeta^2}\r\}
\earr
\ee
The ${\hat E}$ in (4) is the standardized $E$, i.e. it is zero centered and
normalized to unit width, ${\hat E}=(E-\epsilon_H)/\sigma_H$. In the first
step in (4) the fact that EGOE exhibits average fluctuations separation (with
little communication between the two) is used. For example, in the normal
mode decomposition of the EGOE state denity,  it is seen that the long
wavelength parts generate the smoothed Gaussian density (with corrections)
and the short wavelength parts the GOE fluctuations with damping of the
intermediate ones (see \cite{Br-81,Mo-75,Ki-20} for detailed discussions on
this  important result). This allows one to carry out $\l|{\cal C}^E_k\r|^4$
ensemble average independent of the other  smoothed (average) term. In the
second line the Porter-Thomas  form of local strength fluctuations is used
and then $\overline{\l|{\cal C}^E_k\r|^4}$=3, a GOE result. In the third step
the result in (1) is used. Then, the Gaussian forms, valid in the chaotic
domain ($\lambda > \lambda_{F_k}$), of all the densities for EGOE(1+2) give
the final formula (this result was quoted first in \cite{Ko-98} without
details). Before turning to the formula for the localization length $\bl_H$,
let us briefly discuss about the corrections to (4) due to the fluctuations
in the variances of $F_k(E)$; the form with $F_k(E)$ shown  explicitly,  is
written in (4)  for this purpose and this form also allows one to understand
the results in \cite{Kp-00} as discussed ahead.

The correction to NPC due to $\delta \sigma^2_k = 
\sigma_k^2-\overline{\sigma_k^2} \neq 0$ is  obtained by using, for small
$\l| \delta \sigma^2_k \r|$, the hermite polynomial expansion which gives
\cite{Ke-69}, $F_{k:\cg}(E) \rightarrow F_{k:\cg}(E)\l\{1+c_2
(\ce_k^2-1)\r\}$ where  $c_2=\delta\sigma^2_k/2 \overline{\sigma^2_k}$ and
$\ce_k=(E-E_k)/\sqrt{\overline{\sigma^2_k}}$. This corrected $F_k(E)$ is used
in the integral form with $F_k(E)$ in (4).  As NPC involves sum over all the
$\l.\l|k\r.\ran$ states, it is a valid assumption to treat $(\delta
\sigma^2_k)$'s as random  and therefore in $[F_k(E)]^2$ only the terms that
are quadratic in  $(\delta \sigma^2_k)$ will contribute (see \cite{Kp-00}).
Replacing $\l[(\delta\sigma^2_k)/\overline{\sigma^2_k}\r]$ by
$(\delta\sigma^2)/\overline{\sigma^2_k} = \l[(d)^{-1} \{\sum_k\,
(\delta\sigma^2_k)^2\}\r]^{1/2}/\overline{\sigma^2_k}$ and substituting the
corrected $F_k(E)$ for $F_{k:\cg}(E)$ in (4), we get
\be
\barr{rcl}
(\mbox{NPC})_E & = & \dis\frac{(3/d)}{\l[\rho_\cg^H(E)\r]^2} 
\dis\int dE_k \;
\rho_\cg^{H_k}(E_k)\,\l[F_{k:\cg}(E)\r]^2 \l(1+ \dis\frac{(\delta 
\sigma^2)}{2 \overline{\sigma^2_k}} (\ce_k^2 -1)\r)^2  \nn8
& = & \dis\frac{d}{3}\;\dis\sqrt{1-\zeta^4}\;
exp-\l\{\dis\frac{\zeta^2\;{\hat E}^2}{1+\zeta^2}\r\}\;
\l\{1+ \dis\frac{1}{4} \l[\dis\frac{(\delta\sigma^2)}{\sigma^2_H}\r]^2 
X(E)\r\}^{-1}\;;
\earr
\ee
$$
X(E)=\dis\frac{1}{(1+\zeta^2)^4}
\l[\hat{E}^4 -2\dis\frac{(1+\zeta^2)
(1-2\zeta^2)}{1-\zeta^2} \hat{E}^2 + 
\l(\dis\frac{1+\zeta^2}{1-\zeta^2}\r)^2
(1+2\zeta^4)\r]
$$
The $\delta\sigma^2$ corection term in (5) is valid only when the 
fluctuations in the variances of $F_k(E)$'s are small  (this is in general
always true). For small $\zeta$ values, the formula for NPC in (5) reduces to
the expression given recently, for EGOE(2), by Kaplan and Papenbrock
\cite{Kp-00}; they use ideas related to 'scar theory'. In the EGOE(1+2)
hamiltonian $H=h(1) + \lambda V(2)$,  with $\lambda \rightarrow \infty$  one
obtains EGOE(2) and then it is clear from the definition in (2) that in this
limit $\zeta \sim 0$. More precisely, with $N >> m >> 1$,  $\zeta^2 \sim
{\tiny{\l(\barr{c} N \\ 2 \earr \r)}}^{-1}$ and $\l[(\delta \sigma^2)/
\sigma_H^2 \r]^2 \sim \l[{\tiny{\l(\barr{c} m \\ 2 \earr \r)}}
{\tiny{\l(\barr{c} N \\ 2 \earr \r)}}\r]^{-1}$ for $\{H\}=\{V(2)\}$;  see
appendix A. Therefore for finite $N$, the correlation coefficient and the
variance corrections are small but non zero and in the large $N$ limit they
are zero giving the GOE result as pointed out in \cite{Ko-98}. As we add the
mean-field part to the EGOE(2), $\zeta$ increases and at the same time the
variance correction decreases; see Appendix A. Thus the formula (5) with the
($\delta \sigma^2)$ term is  important only for small $\zeta$. Eq. (4) is
accurate for reasonably large  $\zeta$ (say for $\zeta \geq 0.3$) as in the
examples discussed  in \cite{Ko-98}. All these results are well tested by the
numerical examples in Sect. IV.

Proceeding exactly as in (4), formula for the localization length $\bl_H$ as
a function of the excitation energy is derived. Using the definition (3), 
writing $\l|C_k^E\r|^2$ in terms of $\l|{\cal C}^E_k\r|^2$ and 
$\overline{\l|C_k^E\r|^2}$, using the GOE results $\overline{\l|{\cal
C}^E_k\r|^2}$=1 and $\overline{\l|{\cal C}^E_k\r|^2 \;  \ln(\l|{\cal
C}^E_k\r|^2)} =-\ln 0.48$, applying the last equality  in (1) and replacing
all the densities by their corresponding Gaussian forms, converting the sum
in (3) into an integral and finally carrying out the integration, the
expression for $\bl_H$ in wavefunctions is obtained,
\be
\barr{rcl}
\bl_H(E) & \stackrel{EGOE(1+2)}{\longrightarrow} &
-\dis\int dE_k\;\dis\frac{\rho_{biv:\cg}(E,E_k)}{\rho^H_\cg(E)} \ln\l\{
\dis\frac{\rho_{biv:\cg}(E,E_k)}{\rho^{H_k}_\cg(E_k)\,\rho^H_\cg(E)}\r\} \nn8
\\
& = &
\dis\sqrt{1-\zeta^2}\;exp\l(\dis\frac{\zeta^2}{2}\r)\;
exp-\l(\dis\frac{\zeta^2\, {\hat E}^2}{2}\r) 
\earr
\ee
The result in (6) was reported in \cite{Ko-98} without details. By rewriting
the integral in (6) in terms of $F_k(E)$ and making small $(\delta \sigma^2)$
expansion just as in the case of NPC, the formula incorporating corrections 
due to fluctuations (with respect to $k$) in the variances of $F_k(E)$ is
derived following the arguments that led to (5). Neglecting higher order
terms in $\l[(\delta\sigma^2)/\sigma^2_H\r]^2$, the final result is, 
\be
\bl_H(E)=\dis\sqrt{1-\zeta^2}\;exp\l(\dis\frac{\zeta^2}{2}\r)\;
exp-\l(\dis\frac{\zeta^2\, {\hat E}^2}{2}\r) \l(1-\dis\frac{1}{8}
\l[\dis\frac{(\delta\sigma^2)}{\sigma^2_H}\r]^2 Y(E)\r)\;;  
\ee 
$$
Y(E)=\dis\frac{1}{\l(1-\zeta^2\r)^2}\l\{\l(1-\zeta^2\r)^2\, \l({\hat
E}^2-1\r)^2 +4 \zeta^2 \l(1-\zeta^2\r){\hat E}^2 + 2 \zeta^4 \r\} 
$$

\section{Numerical tests}

NPC and $\bl_H$ are calulated for a EGOE(1+2) with 6 particles in 12  single
particle states and the results are compared with (4-7) in  Fig. 2. In the
numerical calculations, the single particle energies
$(i+1/i),\;i=1,2,\ldots,12$ define $h(1)$ (as in \cite{Fl-96} and Fig. 1), in
the two-particle space $V(2)$ is a GOE (calculations use 25 members) with 
unit matrix elements variance and the hamiltonian ensemble is
$\{H_{\alpha,\lambda}\}=\alpha h(1) + \lambda \{V(2)\}$. The value of
$\lambda=0.2$ is fixed so that, for $\alpha \leq 1$ the level fluctuations
are of GOE; i.e. one is in the chaotic domain (see \cite{Ko-00,Ks-00} and
Fig. 1).  Results for $\alpha$ = 0, 0.5 and 1 in Figs. 2a,b  clearly
demonstrate that the   EGOE(1+2) formulas based on the bivariate Gaussian
form for $\rho_{biv}(E,E_k)$ are excellent. In these examples $\zeta$ values
are 0.16, 0.59 and 0.82 respectively. The $(\delta \sigma^2)$ correction is
seen to be important only for the case with $\alpha=0$. In fact the
$\l[(\delta\sigma^2)/\sigma^2_H\r]^2$ values for the three cases considered
are $0.121 \times 10^{-1}$, $0.545 \times  10^{-2}$ and $0.134 \times
10^{-2}$. Thus, for realistic fermion models that are represented by
EGOE(1+2) (with $\lambda > \lambda_{F_k}$),  the correction due to variance
fluctuations is expected to be significant only in the situation $\zeta$ is
small. Extension of EGOE(2) with explicit inclusion of spin degrees of
freedom (each single particle level is taken to be doubly degenerate with
$s_z=\pm \frac{1}{2}$; see the third reference in \cite{Al-97}) was
considered and for a system of 6 fermions in 7 levels (i.e. $m=6$, $N=7
\times 2$) with total $S_z=0$, giving $d= 1225$, NPC was calculated as a
function of the  excitation energy in \cite{Kp-00}; we call this model
EGOE(2)-S. In this example, as given in \cite{Kp-00}, $\zeta = 0.3$ and
$\l[(\delta \sigma^2)/ \sigma^2_H\r]^2 = 0.052$. Thus,  here the corrections
due to variance fluctuations are non-negligible (the situation in this case 
is similar to the $\alpha=0$ case in Fig. 2) and     applying (5) gives
excellent description, as shown in Fig. 3a, of the results for NPC  reported
in  \cite{Kp-00} for the EGOE(2)-S model.  Returning to Fig. 2, it  should be
mentioned that there are differences between the numerical results and the
predictions based on  (4,6) even for the cases with $\zeta$ = 0.59 and 0.82.
These may be due to the departures of $\rho_{biv}(E,E_k)$ from the bivariate
Gaussian form.  An important observation from (4,6) is, at the spectrum
center NPC=$(d/3) \sqrt{1-\zeta^4}$ and
$\bl_H=\sqrt{1-\zeta^2}\;exp(\zeta^2/2)$. Therefore for $\zeta^2$ close to
0.8 or large, there will be large deviations from GOE even at the spectrum
center for a system described by EGOE(1+2). This is clearly seen in the
$\alpha=1$ case in Fig. 2; here $\zeta=0.82$. Finally, it should be mentioned
that the EGOE(1+2) calculations for the $N=14$, $m=7$ system   (the case
considered in Fig. 1) are also carried out and the results are seen to be
essentially same as in Figs. 2a,b.  

Let us now turn to the nuclear shell model which is a realistic interacting 
fermion model. There are shell model results for the $(2s1d)$ shell (here
after called  $sd$ shell) nuclei $^{28}$Si \cite{Ze-96} and $^{22}$Na (see
\cite{Ko-00} and the second reference in \cite{Ko-99}) for NPC and $\bl_H$ in
wavefunctions. For $^{28}$Si the 839 dimensional $J=0,T=0$ space (with six
protons and six neutrons in the $sd$ shell) and the 3243 dimensional
$J=2,T=0$ space are considered. Similarly, for $^{22}$Na the 307 dimensional
$J=2,T=0$ space (with three protons and three neutrons in the $sd$ shell)
is considered. The results for these nuclei are analyzed 
using (4,6) as briefly discussed in \cite{Ko-98,Ko-00}. In all the $sd$ shell
examples, $\zeta \sim 0.6-0.7$ and therefore the situation is similar to the
$\alpha=0.5$ case in Fig. 2.  Thus, in these examples the departures from GOE
at the spectrum center are  no more than 10\% but away from the center, there
are large departures.  The  shell model NPC and $\bl_H$ for $sd$ shell nuclei
are seen to be well  described by the EGOE forms in (4,6). For further
confirming this, NPC is evaluated for $^{24}$Mg in the 325 dimensional
$J=0,T=0$ space (with 4 protons and 4 neutrons in the $sd$ shell) and the
results are shown in Fig. 3b; here $\zeta=0.68$. It can be concluded that the
deviations of the $sd$ shell model results from GOE clearly imply that the
shell model hamiltonians are well represented by EGOE(1+2) (with $\lambda >
\lambda_{F_k}$) but not by GOE. It is also seen that  the corrections due to
$(\delta \sigma^2)$ are small for $(sd)$ shell nuclei (note that here $\zeta$
is large); in the $^{24}$Mg example, $\l[(\delta
\sigma^2)/\sigma_H^2\r]^2=0.024$. In order to further substantiate the  EGOE
description of the structure of nuclear shell model wavefunctions, we have
analyzed using (6)  the $\bl_H(E)$ vs $E$ results reported recently in
\cite{Mo-00} for $2p1f$ shell (hereafter called $pf$  shell) nuclei $^{50}$Ca
and $^{46}$Sc.  In  the case of $^{50}$Ca the 2051 dimensional $J=6,T=5$
space (with 10 protons in the $pf$ shell) and in $^{46}$Sc  the 2042
dimensional $J=1, T=2$ space (with one proton and 5 neutrons in the $pf$
shell) are considered and a modern large shell model code was used  for
obtaining the $\bl_H$ values.  The shell model results for $\bl_H$ in Fig. 9
of \cite{Mo-00}, via (2),  determine $\zeta$ to be 0.96 and 0.92 respectively
for the $^{50}$Ca and $^{46}$Sc examples; results for $^{46}$Sc are shown in
Fig. 3c. From the definition (2) but employing averages over $mT$ spaces
(instead of $mJT$ spaces), we obtain the $\zeta$ values 0.91 and 0.89 
respectively. It should be pointed out that given the single particle
energies and the two-body matrix elements of the shell model hamiltonians, it
is easy to calculate $\zeta$ in fixed $mT$ spaces using trace propagation
methods (by extending (A.3) and (A.4)) \cite{Ko-00,Kk-89}. The $pf$ shell
examples  are similar to the $\alpha=1$ case in Fig. 2 and therefore, as
expected, one sees large departures from GOE even at the spectrum center.
Finally, it is seen from the shell model examples in Fig. 3 and the EGOE
examples in Fig.2 that further corrections to the results in (4-7) need to be
worked out but this is not attempted in this paper. Similarly, study of the
nature of fluctuations in NPC and $\bl_H$ is postponed for future.

\section{Conclusions}

Wavefunction structure given by the EGOE(1+2) random matrix ensemble 
$\{H\}=h(1)+\lambda\,\{V(2)\}$ is studied by deriving compact formulas for
NPC and $\bl_H$. They are based on: (i) the Gaussian form for strength
functions $F_k(E)$'s and the bivariate Gaussian form for 
$\rho_{biv}(E,E_k)$ (with $F_k(E)$ being a conditional density of 
$\rho_{biv}(E,E_k)$) which are valid in the chaotic domain defined by
$\lambda > \lambda_{F_k}$; (ii) there is average-fluctuations separation
(with little communication between the two) in energy levels and strengths  
with local strength fluctuations following the Porter-Thomas law; (iii) there
is a significant unitary group decomposition of the hamiltonian. With
EGOE(1+2), the NPC and $\bl_H$ take Gaussian forms  as a function of the
excitation energy and they are defined by the bivariate correlation
coefficient $\zeta$ which measures the variance of the distribution of
centroids of $F_k(E)$'s relative to the state density variance. Theory for
incorporating corrections due to fluctuations in the variances (with $k$) of
$F_k(E)$ is also given. For small $\zeta$, the present formulation gives back
the results for pure EGOE(2) (i.e. in the $\lambda \rightarrow \infty$
limit of EGOE(1+2)) as derived
in \cite{Kp-00} recently. The formulas derived for NPC and $\bl_H$  are
subjected to numerical EGOE(1+2) tests with $\zeta$ changing from 0.1 to 0.8.
These and the analysis of the results for a EGOE(2)-S example and some 
nuclear shell model examples, clearly point out that isolated finite
realistic interacting particle systems, in the chaotic domain ($\lambda \geq
\lambda_{F_k}$), will have wavefunction  structure  as given by EGOE(1+2). 
Finally, in
the theory given by (4,6), NPC and $\bl_H$ depend on just one parameter and
this appears to be an aspect of 'geometric chaos' (see \cite{Ze-00} for a
recent discussion on 'geometric chaos').

\acknowledgements{
This work has been partially supported by DST(India).}

\appendix
\begin{center}
{\bf Apendix A}
\end{center}

Let us consider a system of $m$ fermions in $N$ single particle states with a
(1+2)-body hamiltonian $H=h(1)+V(2)$ where $h(1)$ is specified by the single
particle energies $\epsilon_i$ (with $i$ denoting the $i^{\mbox{th}}$ single
particle state, $h(1)=\sum_i \epsilon_i\,n_i$ where $n_i$ are number
operators) and $V(2)$ by the two-body matrix elements  
$V_{ijkl}=\lan kl \mid V(2) \mid ij \ran$. The two-body interaction  
can be seperated into $V(2)=V^{[0]}+\bv$ where $V^{[0]}$ is given by 
$$ 
V^{[0]}=\dis\sum_{i<j}
V_{ijij} \, n_i n_j \;\;\;\;\;\;\;\;\;\;\;\;\;\;\;\;\;\;\;\;\;\;\;\;\;
\;\;\;\;\;\;\;\;\;\;\;\;\;\;(A.1) 
$$ 
The $h(1)+V^{[0]}$ generates the $F_k(E)$ centroids $E_k$. With $N$ single
particle states, there is a $U(N)$  group generated by the $N^2$ operators
$a^\dagger_i a_j$ where   $a^\dagger_i$ and $a_j$ are one particle creation
and destruction operators respectively. With respect to this $U(N)$ group, 
$V^{[0]}$ decomposes into $\nu= 0,1,2$ parts and their explicit structure is
(for a given $m$),
$$
\barr{l}
V^{[0],0} = {\tiny{\l(\barr{c} m \\ 2 \earr \r)}}{\tiny{\l(\barr{c} N \\ 
2 \earr \r)}}^{-1} \;
V^0\;\;;\;\;V^0 = \dis\sum_{i<j} V_{ijij} \nn8
V^{[0],1} = \dis\frac{m-1}{N-2} \dis\sum_{i} \zeta_i n_i\;\;;\;\;\zeta_i=
\dis\sum_{j \neq i} \l(V_{ijij} - V^0\r) \nn8
V^{[0],2} = V^{[0]} - V^{[0],0} -
V^{[0],1}\;\;\;\;\;\;\;\;\;\;\;\;\;\;\;\;\;\;\;\;\;\;\;\;\;\;\;\;\;\;\;\;
\;\;\;\;\;\;\;\;\;\;\;(A.2)
\earr
$$
Similarly the $h(1)$ operator will have $\nu=0,1$ parts;  $h^0=m \epsilon^0$
where $\epsilon^0=(N)^{-1}\sum_i \epsilon_i$ and $h^1(1) =
\dis\sum_i\,\epsilon^1_i n_i$ where $\epsilon^1_i= \epsilon_i-\epsilon^0$.
Finally it is to be noted that $\bv$ behaves as a $\nu=2$ operator.

The $U(N)$ norm (in the $m$-particle spaces) 
of an operator $\co$ is defined by $\l|\l|\co\r|\r|_m=
\sqrt{\lan (\co-\lan \co \ran^m)^\dagger (\co-\lan \co \ran^m) \ran^m}$. 
An important theorem is that the $\nu=0,1,2$ parts
of $H$ are orthogonal with respect to this $U(N)$ norm. For a $\nu=1$ 
operator $\co(1)=\sum_i e_i n_i$, the norm square is simply given by 
$$
\l|\l|\co(1)\r|\r|^2_m=\dis\frac{m(N-m)}{N(N-1)} \dis\sum_i e_i^2
\;\;\;\;\;\;\;\;\;\;\;\;\;\;\;\;\;\;\;\;\;\;\;\;\;\;\;\;\;\;\;\;\;\;\;
\;\;\;\;\;\;\;\;\;\;\;(A.3)
$$
Similarly for a $\nu=2$ operator $\co(2)$,
$$
\l|\l|\co(2)\r|\r|^2_m=\dis\frac{m(m-1)(N-m)(N-m-1)}{2(N-2)(N-3)} 
\lan\cod(2)\co(2)\ran^2
\;\;\;\;\;\;\;\;\;\;\;\;\;\;\;\;\;\;\;\;\;(A.4)
$$
Using (A.3) and (A.4) one can calculate the norms of $h^1+V^{[0],1}$ and
$V^{[0],2}$ and in general the later is very small compared to the former.
Then $h(1)+V^{[0]} \rightarrow \bh = \sum_i\,\xi_i n_i$ where 
$\xi_i=\epsilon^1_i + \frac{m-1}{N-2} \zeta_i$ (note that at the end we add
the  spectrum centroid  generating part $h^0+V^{[0],0}$ to $\bh$). Thus,
neglecting the $V^{[0],2}$ part, the centroids of $F_k(E)$'s are generated by
$\bh$ and the variances by $\bv$. As $\bh$ and $\bv$ are orthogonal,
$\sigma^2_H=\sigma^2_\bh + \sigma^2_\bv$. These variances, in $m$-particle
spaces, follow easily from (A.3,A.4). See \cite{Ko-00,Fr-71} for  further
details.

Let us consider a EGOE(1+2) hamiltonian $H=\alpha h(1) + \lambda V(2)$  with
unit spacing between the $\epsilon_i$'s and the $V_{ijkl}$ taken as
zero centered Gaussian variables with  unit variance. In the $N >> m >>1$
situation one can study the behaviour of $\zeta^2$ and $(\delta \sigma^2)$ as
follows. The correlation coefficient $\zeta^2 = \sigma^2_\bh/\sigma_H^2$ and,
neglecting the contributions of $V(2)$ to $\sigma_\bh$, one gets
$\sigma^2_\bh \sim (m N^2/12)\alpha^2$. Similarly $\sigma^2_\bv \sim
{\tiny{\l(\barr{c} m \\ 2 \earr \r) \l(\barr{c} N \\ 2  \earr
\r)}}\lambda^2$. Here (A.3) and (A.4) are used. Therefore
$\zeta^2=\l[(1+3m\,(\lambda/\alpha)^2\r]^{-1}$ and this expression gives 0.51
and 0.76 for the $\alpha=0.5$ and 1 cases in Fig. 2. They compare well
with the exact numbers given in Fig. 2. However this estimate fails in the
situation $\alpha \rightarrow 0$. For $\alpha=0$ the $\bh$ has to be replaced
by $V^{[0]}$ and then the $E_k$ energies are a sum of  ${\tiny{\l(\barr{c} m
\\ 2 \earr \r)}}$ zero centered Gaussian variables  each with variance
$\lambda^2$. This together with the $\sigma^2_\bv$ expression, gives $\zeta^2
\sim  {\tiny{\l(\barr{c} N \\ 2 \earr \r)}}^{-1}$ for $\alpha \sim 0$ as 
pointed out in \cite{Kp-00}. The number quoted for the $\alpha=0$ case in
Fig. 2 is close to this estimate. Finally an estimate for  $[(\delta
\sigma^2)/\sigma^2_H]^2$ is obtained from (A.4) by noting that $\sigma_\bv^2$
is a sum of $K \sim {\tiny{\l(\barr{c} m \\ 2 \earr \r)}}{\tiny{\l(\barr{c} N
\\ 2  \earr \r)}}$  $\chi^2$-variables and therefore $[(\delta
\sigma^2)/\sigma^2_\bv]^2 = 2/K$ as given first in \cite{Fl-96}. Then,
$\sigma^2_\bv=(1-\zeta^2)\sigma^2_H$  gives the final result $[(\delta
\sigma^2)/\sigma^2_H]^2 \sim  2(1-\zeta^2)/{\tiny{\l(\barr{c} m \\ 2 \earr
\r)}}{\tiny{\l(\barr{c} N \\ 2 \earr \r)}}$.

\newpage
\begin{center}
{\bf FIGURE CAPTIONS}
\end{center}

\noindent {\bf Fig. 1} Strength functions  $F_k(E)$,  Dyson-Mehta 
$\overline{\Delta}_3$ statistic for level fluctuations and occupancies 
$\lan E \mid n_i \mid E \ran$ for EGOE(1+2) for various
values of the interaction strength $\lambda$ in $\{H\} = h(1) + \lambda
\{V(2)\}$ for a system of 7 fermions in 14 single particle states (due to
computational constraints, here only one member is considered just as in
\protect\cite{Fl-96}); the matrix dimension is 3432. The single particle
energies used in the calculations are $\epsilon_i=(i+1/i), i=1,2,\ldots,14$
just as in \protect\cite{Fl-96}. (a) the histograms are EGOE(1+2) results for
the strength functions,  continuous curves are BW fit and the dotted curves
are Gaussian for $\lambda \leq 0.1$ and Edgeworth corrected Gaussian 
\cite{Ke-69} for $\lambda > 0.1$. In constructing the strength functions,
$|C_k^E |^2$ are summed over the basis states $\l.\l| k \r. \ran$ in the
energy window ${\hat{E}}_k \pm \Delta$ and then the ensemble averaged
$F_{{\hat{E}}_k} ({\hat{E}})$ vs ${\hat{E}}$ is constructed as a histogram;
the value of $\Delta$ is chosen to be 0.025 for $\lambda \leq 0.1$ and beyond
this $\Delta=0.1$. Here ${\hat{E}}_k = (E_k-\epsilon_H)/\sigma_H$ and in the
figure ${\hat{E}}_k=0$.  Note that for $\lambda_{F_k} \sim 0.2$ there is BW
to Gaussian transition. (b) The ${\overline{\Delta}}_3(L)$ statistic for
overlapping intervals of  length $L \leq 40$ are compared with Poisson and
GOE values. For $\lambda \sim 0.06$, there is Poisson to GOE transition in
the $\overline{\Delta}_3$ statistic.  (c) The wavy curves are 
numerical EGOE(1+2)
results for occupancies and the smoothed curves  with $\lambda \geq 0.06$
correspond to the results of EGOE(2) theory (ratio of Gaussians). Note that
for $\lambda < 0.06$ there are wide fluctuations in occupancies and the
smoothed forms here are meaningless. All the results are shown for the lowest
6 single particle states. Results similar to those in  the figure, for the
$N=12,m=6$ case are reported in \protect\cite{Ko-00}.

\noindent {\bf Fig. 2} (a) Number of principal components NPC and (b) the
localization length $\bl_H$ in wavefunctions  for a system of 6 interacting
particles in 12 single particle states (matrix dimension is 924).  Here, for
conveniance, the EGOE(1+2) hamiltonian is changed to
$\{H_{(\alpha,\lambda)}\}=\alpha h(1) + \lambda \{V(2)\}$. Numerical
EGOE(1+2) results correspond to filled circles. The continuous curves
correspond to the theory (4) for NPC and (6) for $\bl_H$. For the case with 
$\alpha=0$, the dashed curves correspond to the theory (5) for NPC and (7)
for $\bl_H$.  For other cases the correction due to variance fluctuations is
negligible and hence only the results of (4,6) are shown in the figure. Note
that NPC=$d/3$ and $\bl_H=1$ for GOE. See text for further details.

\noindent {\bf Fig. 3} (a) NPC for the EGOE(2)-S model described in the text
compared with the results given by (4,5). The filled circles are for the
numerical EGOE(2)-S calculations reported in \cite{Kp-00}. The continuous and
dashed curves represent (4) and (5) respectively. (b) NPC for the $(sd)$
shell nucleus $^{24}$Mg compared with (4). The shell model calculations are
same as in \cite{Ko-99}. In this example (4) and (5) give almost identical
results and hence the curve corresponding to (5) is not shown in the figure.
(c) Shell model results for $^{46}$Sc for $\bl_H$ reported in \cite{Mo-00}
compared with the theoretical curve given by (6) with $\zeta=0.92$.

\ed